\documentclass{article}
\usepackage{ijcai23}

\usepackage{times}
\usepackage{soul}
\usepackage{url}
\usepackage[hidelinks]{hyperref}
\usepackage[utf8]{inputenc}
\usepackage[small]{caption}
\usepackage{graphicx}
\usepackage{amsmath}
\usepackage{amsthm}
\usepackage{booktabs}
\usepackage{algorithm}
\usepackage{algorithmic}
\usepackage[switch]{lineno}

\title{From Words to Music: A Study of Subword Tokenization Techniques in Symbolic Music Generation}

\author{
Adarsh Kumar$^1$
and
Pedro Sarmento$^2$
\affiliations
$^1$Indian Institute of Technology Kharagpur, India\\
$^2$ Queen Mary University of London, UK\\
\emails
adarshkumar712.ak@gmail.com,
p.p.sarmento@qmul.ac.uk
}

\begin{document}

\maketitle

\begin{abstract}
Subword tokenization has been widely successful in text-based natural language processing (NLP) tasks with Transformer-based models. As Transformer models become increasingly popular in symbolic music-related studies, it is imperative to investigate the efficacy of subword tokenization in the symbolic music domain. In this paper, we explore subword tokenization techniques, such as byte-pair encoding (BPE), in symbolic music generation and its impact on the overall structure of generated songs. Our experiments are based on three types of MIDI datasets: single track-melody only, multi-track with a single instrument, and multi-track and multi-instrument. We apply subword tokenization on post-musical tokenization schemes and find that it enables the generation of longer songs at the same time and improves the overall structure of the generated music in terms of objective metrics like structure indicator (SI), Pitch Class Entropy, etc. We also compare two subword tokenization methods, BPE and Unigram, and observe that both methods lead to consistent improvements. Our study suggests that subword tokenization is a promising technique for symbolic music generation and may have broader implications for music composition, particularly in cases involving complex data such as multi-track songs.
\end{abstract}


\section{Introduction}\label{sec:Intro}
Subword tokenization is a widely used technique for text representation in natural language processing (NLP). Such tokenization techniques based on the creation of subword tokens, like byte pair enconding (BPE) \cite{sennrich-etal-2016-neural}, Unigram \cite{kudo-2018-subword} and WordPiece \cite{WordPiece}, have become ubiquitous in various NLP tasks. Attributed to their efficiency in modeling longer patterns, rather than simply characters, these subword tokenization techniques became extremely successful with Transformer models like BERT \cite{BERT} and GPT \cite{GPT}, achieving state-of-the-results in multiple text-based NLP applications. Works like \cite{park-etal-2020-empirical} and \cite{galle-2019-investigating} have further shown the universality of its application across languages, not just in English.

\begin{figure}[t]
\centering
\includegraphics[width=1.0\columnwidth]{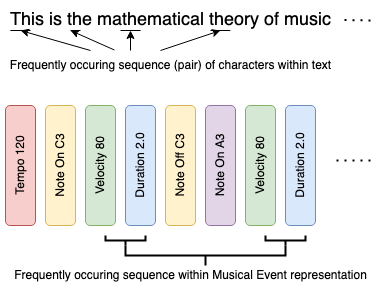}
\caption{Example of similar frequently co-occurring structures within the text (at character level) and musical representations (MIDI-like, at event level).}
\label{fig:example_subword}
\end{figure}

Inspired by the success of Transformer models in text-NLP,\footnote{For the sake of clarity and distinction, we will henceforth refer to tasks related to the text in NLP as text-NLP.} recent years have witnessed a shift in research work towards leveraging Transformers \cite{Vaswani2017} in the domain of symbolic music generation \cite{MusicTransformer,transformer-gan,Infil}. This shift can be ascribed to the resemblance of symbolic music post-musical-tokenization\footnote{Musical Tokenization refers to tokenization of symbolic formats like MIDI or GuitarPro with Music Tokenization such as REMI or MIDI-like.} to that of text tokens. With Transformers' extraordinary capability to model longer sequences, we are able to generate coherent, adequate pieces of music end-to-end \cite{Hsiao2021,Huang2020,SentTransformerGAN}. 

Despite all the success of the predecessors in improving the state of music generation, these models are often accused of failing to entirely capture the repetitive structure and overall musical development of songs \cite{failmusicrepete,Jazz,composenembel}. This becomes more apparent as the structure of the music becomes complex such as in the case of polyphonic music or multi-track music. A reasonable explanation for this could be the significantly longer sequence of symbolic music tokens, which limits the segment of a song visible to the Transformer model, hindering its understanding of the overall musical structure. As an analogy, this would be equivalent to representing a piece of text as a sequence of individual characters. 

One possible solution to this problem is at the token level. The idea is to group individual events into subgroups, similar to subwords in text-NLP. There have been works like \cite{Hsiao2021} and \cite{musicBPE}, which tried to group musical events, exploiting the properties of musical structure for certain MIDI-based datasets. However, most of these works are dependent on the musical structure of certain datasets involved, and hence cannot be extrapolated to other formats easily. So, to the best of our knowledge, no work has been done so far to assess the use of subword tokenization techniques like BPE and Unigram that utilize the co-occurrence-based structure of musical events, independent of the musical structure of the training dataset itself.

We are therefore motivated to study whether the use of subword tokenization can improve the overall musical structure of generated songs, while at the same time being independent of the dataset or format of symbolic music involved. In this work, we specifically investigate the usefulness of subword tokenization techniques like BPE \cite{sennrich-etal-2016-neural} and Unigram \cite{kudo-2018-subword} in modeling the task of symbolic music generation. Through our experiments, we try to answer the following two primary questions:

\begin{itemize}
    \item Q1: Can we use subword tokenization techniques to improve the overall musical structure and musical quality of the generated examples?
    \item Q2: How do these findings generalize between two different subword tokenization techniques, namely BPE and Unigram?
\end{itemize}

\noindent In an effort to answer the above questions, our main contributions through this paper are as follows:
\begin{itemize}
    \item Creation and implementation of an evaluation environment to objectively assess the usefulness of subword tokenization techniques, namely BPE and Unigram, to improve overall musical structure (in terms of quantitative metrics like Pitch Class Entropy, Structureness Indicators etc.), independent of data-specific factors like music file formats (e.g. MIDI or GuitarPro);
    \item Establishing the efficiency of subword tokenization across datasets, data formats and musical tokenization techniques, with our study involving melody-only, polyphonic and multi-track datasets;
    \item Demonstrating the usefulness of subword tokens towards facilitating longer pieces of generated music within the same inference time.
\end{itemize}


\section{Background and Related Work}

\subsection{Subword Tokenization}
Tokenization has become a fundamental process in NLP which involves breaking the macroscopic units of text such as `words' or `sentences' into smaller units called tokens. Since representing the much longer textual data as individual characters is highly ineffective, the concept of `subword tokenization' was introduced \cite{Subword}. It involves breaking the word or sentence into sub-words (a subsequence of characters with length $\geq 1$), which are then used to represent the data. Over the years, several subword tokenization techniques have been introduced such as BPE \cite{Gage,sennrich-etal-2016-neural}, Unigram \cite{kudo-2018-subword}, WordPiece \cite{WordPiece} and SentencePiece \cite{SentencePiece}. Most of these techniques involve the selection of subwords based on their frequency within the training data. 

Byte-Pair-Encoding (BPE) \cite{sennrich-etal-2016-neural} is a data compression technique \cite{Gage} which involves the replacement of the most frequent pair of bytes by a single, unused byte. In text-NLP, BPE is applied to subword units, rather than bytes, by finding the most frequent character n-grams in a text corpus and merging them into a single token. This allows the model to learn a more fine-grained representation of the language, especially for rare or out-of-vocabulary words. The result is a variable-length subword vocabulary, which balances the ability to capture complex language structure with reducing the risk of overfitting.

In contrast to BPE, Unigram \cite{kudo-2018-subword} is another subword tokenization technique, which starts from a large vocabulary and gradually trims the vocabulary towards a smaller one. It involves a probabilistic Unigram language model, which decides whether to keep a subword or not based on its likelihood and loss function. Both these models have been extensively used in various NLP applications such as \cite{Wang2020}, \cite{Roberta} and \cite{Neurips}. Furthermore, papers like \cite{galle-2019-investigating} and \cite{park-etal-2020-empirical} have shown that these subword tokenization techniques can be applied effectively across languages. This motivates us to explore whether these results can be extended to symbolic music as a language, and what impact it could have on the overall musical structure of generated songs. 

\begin{figure*}[t]
\centering
\includegraphics[width=1\textwidth]{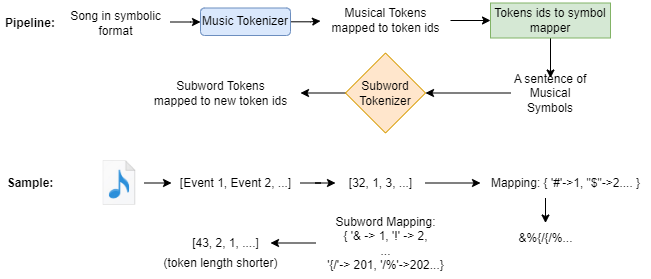}
\caption{Symbolic music processing pipeline with subword tokenization. Please note that, here, musical symbols refer to mapping of musical events to unicode symbols or characters, which are then processed with subword tokenizer.}
\label{fig:preprocessing}
\end{figure*}

\subsection{Symbolic Music Generation}
Symbolic Music Generation involves representation of music data from formats like MIDI or GuitarPro with symbols or sequences of events which are then used to train generative models. This has been an extensively researched area, where researchers are trying to come up with algorithms and models that can generate music at par with human performance \cite{Hiller}. Recently, the domain of symbolic music generation has witnessed steady improvements, mostly driven by advances in deep learning architectures. Overall, approaches towards symbolic music generation with deep learning can be aggregated according to the architecture used, namely Variational Autoencoder (VAEs) models \cite{Tan2020}, Generative Adversarial Networks (GANs) \cite{Dong2018}, and models that stem from natural language processing (NLP) field, such as Recurrent Neural Networks (RNNs) \cite{Meade2019}, Long Short-Term Memory (LSTMs) \cite{Sturm2016}, or Transformers \cite{Vaswani2017}. As stated before, the Transformer architecture \cite{Vaswani2017} is suitable for generating longer sequences when compared to previous approaches used by RNNs. Transformer-like sequence-to-sequence models are able to learn the dependencies and patterns among elements of a given sequence. The work by \cite{AnnaHuang2019}, the Music Transformer, pioneered the application of the self-attention mechanism to generate longer sequences of symbolic piano music. Other examples include works like Musenet \cite{christine_2019}, in which a large-scale Transformer model, GPT-2, was used to generate symbolic multi-instrument music from different musical genres, the Pop Music Transformer \cite{Huang2020}, which uses Transformer-XL \cite{Dai} as a backbone architecture and is able to generate pop piano symbolic music with a better rhythmic structure, the Compound Word Transformer \cite{Hsiao2021}, that presents novel and more efficient ways of tokenizing symbolic music for training purposes, and GTR-CTRL \cite{gtrctrl} explores genre and instrument conditioning using special control tokens with multi-track music dataset DadaGP, thereby controlling overall musical structure of generated songs.

\section{Methodology}
In order to evaluate the usefulness of subword tokenization and to answer the questions mentioned in Section \ref{sec:Intro}, we designed an empirical study where we objectively assess the performance of models with and without subword tokenization methods, while keeping all the other factors identical. For clarification, we here refer to the model without subword tokens as the `base model', while the other models are named according to the respective subword tokenization technique used while modeling (e.g. `BPE model' as the model that used BPE for subword tokenization).

\subsection{Datasets and Music Tokenization Schemes} 
As mentioned earlier, we used three datasets for our experiments, namely the Folk Songs dataset \cite{Sturm2016}, the MAESTRO dataset\cite{MAESTRO}, and the DadaGP dataset \cite{Sarmento2021}. Furthermore, for the music tokenization procedure, we utilized REMI \cite{Huang2020}, Midi-like \cite{midilike2018} and DadaGP tokenization \cite{Sarmento2021} respectively, primarily to demonstrate the compatibility of subword tokenization with existent different music tokenization approaches. Given that training the models such as Music Transformer \cite{MusicTransformer} and Transformer-XL \cite{Tr-XL,Huang2020} on such large datasets is very resource intensive, we restricted our study to subsets of each dataset, using \textbf{1000} samples from the Folk Songs dataset, \textbf{400} samples from the MAESTRO dataset and \textbf{2000} songs from the Rock genre in DadaGP. However, since similar settings were kept for both the models with and without subword tokenization, our restriction does not impact any conclusions from our results.

\subsection{Data Processing}

While there are several subword tokenization methods available, we focused our experiments on two of the most common ones, BPE and Unigram, primarily because of their ease of applicability to any dataset. To leverage subword tokenization with symbolic music, we first converted songs from symbolic format (MIDI or GuitarPro) to musical events using music tokenization schemes such as REMI \cite{Huang2020}, following which we create a mapping, from musical events to unicode symbols. Thus, we obtained a relatively easy-to-process corpus of symbols, in which we assume each song is a single entity similar to SentencePiece \cite{SentencePiece}. We then trained subword tokenization methods on the respective datasets to create a larger vocabulary of subword tokens. Furthermore, we used this vocabulary to process musical tokens of all songs in a given dataset. A statistical summary of the original vocabulary size of musical tokens and subword tokens used in these experiments is given in Table \ref{tab:vocab}.

\begin{table}
    \centering
    \begin{tabular}{l||r|r|r}
        \toprule
        Music Tokenization & Original & BPE & Unigram \\
        \midrule
        REMI   & 227    & 300   & 300   \\
        MIDI-Like & 331    & 1000  & 1000  \\
        DadaGP    & 2104   & 5000  & 5000      \\
        \bottomrule
    \end{tabular}
    \caption{Vocabulary size of original tokens (using respective musical tokenization), post-processing with BPE and Unigram.}
    \label{tab:vocab}
\end{table}


\subsection{Experimental Configuration}

As stated before, in order to assess the independence of results in terms of improvements while using subword tokenization methods against factors like type of dataset, music tokenization procedure, and model, we experimented with three different combinations:

\begin{enumerate}
    \item Folk Songs (monophonic, single instrument, MIDI format) + REMI tokenization + Music Transformer;
    \item MAESTRO (polyphonic, single instrument, MIDI format) + MIDI-Like tokenization + Music Transformer;
    \item DadaGP (polyphonic, multi-instrument, GuitarPro format) + DadaGP tokenization + Transformer-XL;
\end{enumerate}


The key idea here is to experiment with subword tokenization schemes in different settings and configurations and to see if the results in terms of improvement hold true irrespective of the rest of the factors. Furthermore, the choice of monophonic, polyphonic, and multi-track music, allows us to evaluate the usefulness of subword tokenization for music generation tasks with different levels of complexities.


\section{Evaluation Metrics}
To objectively evaluate the results of applying subword tokenization in music generation, we subdivided dedicated metrics into categories: musical quality and structure and efficiency in representation.
\subsection{Musical Quality and Structure}
\textbf{Structureness Indicator (\textit{SI}):} Proposed in \cite{Jazz},
the structureness indicator (SI) is designed to capture the structureness of music, induced by repetitions of musical content. It is based on the fitness-scapeplot algorithm \cite{scape} and the self-similarity matrix (SSM) \cite{ssm} computed from \textit{fitness}, where the degree of repetition is derived from SSM for a given segment ($i$, $j$). Similar to \cite{Jazz}, in our experiments, we used $SI_3^8$, $SI_8^{15}$ and $SI_{15}$, in which $SI_l^u$, $l$ and $u$ represent the lower and upper bound of intervals of consideration (in seconds). We respectively refer to these as \textit{SI-short}, \textit{SI-medium}, and \textit{SI-long}, as they are used to examine the short, medium, and long-term structureness of generated songs. Here it is important to note that a higher \textit{SI-x} value doesn't necessarily mean `better' music, since the generated samples may then be too repetitive. Instead, a more valid assumption is that it is better the closer it is to the real music (i.e. testing corpus), which will be the basis of our evaluation.

\vspace{1mm}
\noindent\textbf{Pitch Class Entropy ($\mathcal{H}$):} Also described in \cite{Jazz}, gives an insight about different pitches, and thereby tonality used in a song. Here, the main idea is to calculate entropy from a normalized 12-dimensional pitch class histogram (corresponding to 12 pitch classes C, C\#, D, ... B) and analyze how close these values are to the real values.

\vspace{1mm}
\noindent\textbf{Groove Pattern Similarity ($\mathcal{GS}$):} Another metric defined in \cite{Jazz}, Groove Pattern Similarity helps in measuring the rhythmic consistency within a song. It calculates the pairwise similarity of each bar's groove vector \textbf{g} (indicating the position in bar with at least a note onset) as $1 - HammingDist(g_a, g_b)$, across all pairs $g_a$ and $g_b$. Similar to the other two metrics mentioned previously, in this case also, the closer the values of groove similarity from the generated songs are to the real songs, the better. 

\subsection{Efficiency in Representation}
\noindent\textbf{Average Number of Tokens per Song:} Within this metric we measure the average number of tokens per song present post-data processing, which are then fed into a Transformer model for training. A more efficient representation will be the one, which has smaller average number of tokens per song.

\vspace{1mm}
\noindent\textbf{Average Number of Tokens per Song for the Same Inference Time:} In this metric, for a given dataset, we generate an equal number of tokens (i.e. same inference)for each of the three models we experimented with. After a post-conversion process back to the original musical tokens, we compare the average number of tokens generated in terms of the original tokens. This metric helps us assess how efficient the representation is in terms of generating longer music within the same inference time. Hence, for this particular metric, the larger the value, the better representation of data will be.

\subsection{Other Metrics} 
\label{sec:othermetrics}
\textbf{NLL Loss:} Negative log-likelihood is a common metric, often used to measure how well a model fits the training dataset \cite{MusicTransformer,nll_metrics,Hsiao2021}. While a relatively closer and smaller value of NLL represents a well-fitted model, in our case we observe that it is not a good metric to compare performances across models, since the model with subword tokenization has a higher number of parameters. Furthermore, a lower NLL loss doesn’t necessarily mean better generation quality. We have still added this to give an idea of how well the BPE or the Unigram model fits relative to the base model.



\section{Results}

\subsection{Experimental Settings}
The experiments were conducted using HuggingFace \cite{wolf2020huggingfaces} and PyTorch implementations of Music Transformer\footnote{\url{https://github.com/jason9693/MusicTransformer-pytorch}} and  Transformer-XL\footnote{\url{https://github.com/YatingMusic/compound-word-transformer}}. For musical tokenization, we used the MidiTok \cite{Miditok} library, which we further processed for subword tokenization using Huggingface's tokenizers\footnote{url{https://huggingface.co/docs/tokenizers/index}}. For the first two parts of our experiment (i.e. Folk Songs and MAESTRO songs dataset involving the Music Transformer model), we used a 3-layer Transformer architecture, with an embedding dimension 256, which we trained on Google Colab Free Tier with a P100 16GB GPU machine. For the last part (i.e. DadaGP with Transformer-XL), we used the same architecture as in \cite{Huang2020}, training the model on a 24GB Quadro RTX 6000 GPU. For the evaluation, we used implementation of the metrics described in the previous section in MusPy \cite{Muspy} and MusDr\footnote{\url{https://github.com/slSeanWU/MusDr}}. Finally, we evaluated the model performance by generating 20 songs for each model configuration. Samples from the generation can be accessed \href{https://drive.google.com/drive/folders/1SQZ422-27kAl3zv65mqZG0bxUvHLbWM5?usp=share_link}{here}.


\subsection{Objective Evaluation}

Results from the three distinct experiments can be seen in tables \ref{tab:folk}, \ref{tab:maestro} and \ref{tab:dadagp}.

\label{sec:objeval}
\begin{table}[h!]
    \centering
    \begin{tabular}{l||c||c|c|c}
        \toprule
        Metrics & Real & Original & BPE & Unigram \\
        \midrule
        SI-short        & 0.4637 & 0.2707 & \textbf{0.3376} & 0.2712  \\
        SI-medium       & 0.4959 & 0.2759  & \textbf{0.3379} & 0.2719  \\
        SI-long         & 0.4543 & 0.2583  & \textbf{0.3340} & 0.2451  \\
        $\mathcal{H}$   & 2.6011 & 2.6924  & \textbf{2.6754} & 2.6842  \\
        $\mathcal{GS}$  & 0.9987 & 0.9984  & \textbf{0.9986} & 0.9985  \\
        \bottomrule
    \end{tabular}
    \caption{Results with the Folk Songs dataset.}
    \label{tab:folk}
\end{table}

\begin{table}[h!]
    \centering
    \begin{tabular}{l||c||c|c|c}
        \toprule
        Metrics & Real & Original & BPE & Unigram \\
        \midrule
        SI-short        & 0.3228 & 0.5119  & 0.3880 & \textbf{0.3483}  \\
        SI-medium       & 0.3066 & 0.4663  & 0.3334 & \textbf{0.2828}  \\
        SI-long         & 0.2343 & 0.4173  & 0.3031 & \textbf{0.2205}  \\
        $\mathcal{H}$   & 3.0555 & 2.5152  & 2.8705 & \textbf{2.9297} \\
        $\mathcal{GS}$  & 0.9917 & 0.9971  & 0.9942 & \textbf{0.9936}  \\
        \bottomrule
    \end{tabular}
    \caption{Results with the MAESTRO songs dataset.}
    \label{tab:maestro}
\end{table}

\begin{table}[h!]
    \centering
    \begin{tabular}{l||c||c|c|c}
        \toprule
        Metrics & Real & Base & BPE & Unigram \\
        \midrule
        SI-short        & 0.5069 & \textbf{0.5110} & 0.5125 & 0.4894 \\
        SI-medium       & 0.5270 & 0.4219 & \textbf{0.4892} & 0.4539 \\
        SI-long         & 0.4972 & 0.2943 & \textbf{0.4397} & 0.3924 \\
        $\mathcal{H}$   & 2.5842 & 2.0529 & 2.3084 & \textbf{2.4333} \\
        $\mathcal{GS}$  & 0.9991 & 0.9994 & 0.9993 & \textbf{0.9992} \\
        \bottomrule
    \end{tabular}
    \caption{Results with the DadaGP dataset.}
    \label{tab:dadagp}
\end{table}

As can be observed from the tables, the use of subword tokenization methods outperforms the base models by a significant margin in both cases (i.e. BPE and Unigram), in almost all the configurations, irrespective of the model, dataset, data format, or music tokenization scheme used.  The values of SI-short, SI-medium, and SI-long, which closely resemble real song data, indicate an overall improvement in the musical structure of songs using subword tokenization. Additionally, longer repetitive structures exhibit more significant improvements compared to shorter ones. The results suggest that subword tokenization techniques have the potential to model the musical structure better and can leverage the latent co-occurring structure within musical tokens to improve the quality of generated music.

That being said, the relatively smaller improvements in the case of the Folk Songs dataset, suggest a correlation between the complexity of the dataset being modeled and the performance change, with more scope for improvements in the case of datasets with higher complexities such as MAESTRO and DadaGP. Adding to this, negligible improvements in terms of SI for Unigram could be possible because the subword tokens generated with Unigram are harder to learn by the model, thereby collapsing to simpler tokens (which is a case similar to base model, which also works on simpler tokens). This again could be possibly attributed to the relative simplicity in terms of structure and lesser scope of co-occurrence when the dataset is melody-only. However, when we move to more complicated datasets, we have musical co-occurring structures like Chords, which improve the feasibility of using Subword Tokenization Techniques.    

\begin{table}
    \centering
    \begin{tabular}{l|c|c|c}
        \toprule
        Dataset & Base & BPE & Unigram \\
        \midrule
        FOLK      & 796 & \textbf{359} & 436 \\
        MAESTRO   & 12925 & \textbf{8831} & 8618 \\
        DadaGP    & 5332 & \textbf{2875} & 2954 \\
        \bottomrule
    \end{tabular}
    \caption{Average number of tokens per song in each representation.}
    \label{tab:AvgTokens}
\end{table}

\begin{table}[h]
    \centering
    \begin{tabular}{l|c|c|c}
        \toprule
        Dataset & Base & BPE & Unigram \\
        \midrule
        FOLK      & 500 & \textbf{1307} & 994 \\
        MAESTRO   & 1000 & \textbf{1570} & 1534 \\
        DadaGP    & 1000 & 1437 & \textbf{1828 }\\
        \bottomrule
    \end{tabular}
    \caption{Average number of tokens generated for same inference time in a dataset (i.e. for same time taken to generate $x$ base model (without subword tokenization) tokens, corresponding $y$ and $z$ tokens generated (in terms of original musical tokens) for BPE and Unigram models respectively.)}
    \label{tab:AvgTokensInf}
\end{table}

Furthermore, the figures from Tables \ref{tab:AvgTokens} and \ref{tab:AvgTokensInf} demonstrate the improved efficiency of representation with the use of subword tokenization. This adds up to the advantage of using subword tokenization with the base model, as it allows to model of longer sequences within the same inference time from a model. These results become particularly important in the case of complex symbolic music datasets such as DadaGP. This dataset, being complex in representation, requires that much longer sequences are generated even for a short song segment. Using subword tokenization with such datasets can allow shortening the sequences, thereby allowing longer music to be inferred. 

A comparison between the performance of models with BPE or Unigram tokenization throughout the results suggests that the improvements hold true in general for subword tokenization, leveraging the frequent cooccurrence of musical events within songs. Though there are some localized differences on what method is more efficient in terms of structural modeling or musical quality, with one working better than the other in certain cases or datasets, in all, these perform better than the base model `without' subword tokenization. This answers our second question of how the results of our study generalize for two different subword tokenization methodologies. Furthermore, this generalization is independent of the constraints of a given musical structure of a particular dataset.

Another interesting aspect of our results is to observe the improvement of results with the experiments involving Transformer-XL i.e. beyond fixed length modeling. While the main purpose of using this model, as proposed in \cite{Huang2020} for music generation, is to model musical structure beyond a fixed length of the input, it is intriguing to see the improvements with subword tokenization even in this case, particularly with longer repetitive structureness. This suggests while the dedicated architecture of Transformer-XL is capable of modeling shorter musical structure better, there is still a loss of information as the model propagates through the windows of Transformer-XL input, leading to lesser long-term repetitiveness. However, on reducing this length with subword tokenization, this loss can be reduced, thereby allowing improved representation of the musical structure, closer to the real data, we are trying to model.


\begin{table}
    \centering
    \resizebox{\columnwidth}{!}{%
    \begin{tabular}{l|l|l|c|c|c}
        \toprule
        Dataset & Model & Train Time & Base & BPE & Unigram \\
        \midrule
        FOLK    & MT & $\sim20 mins$ & 0.08 & 0.13 & 0.11 \\
        MAESTRO & MT & $\sim3 hrs$ & 2.58 & 3.16 & 3.62 \\
        DadaGP  & Tr-XL & $\sim3 days$ & 0.10 & 0.11 & 0.11 \\
        \bottomrule
    \end{tabular}%
    }
    \caption{Ngeative Log-Likelihood Loss for the models (MT $\to$ Music Transformer, Tr-XL $\to$ Transformer XL). As design choices, here we decided the train time based on the complexity of the dataset and the model in use, having kept the same time for all three experiments.}
    \label{tab:NLL}
\end{table}

Lastly, it is important to note the NLL loss performance of the models we trained from Table \ref{tab:NLL}. While the closer values of loss functions in the case of the base, BPE and Unigram suggest the model is able to model the dataset almost equivalently in all the cases, differences in the objective evaluation of generated musical quality, indicate that not all the information is captured within NLL loss. Furthermore, it supports our initial assumption (in section \ref{sec:othermetrics})that NLL is not an appropriate metric to act upon conclusively over the model performances. However, it can still provide an overview of how well the model is fitting the dataset, which in the case of usage of subword tokenization is almost the same as without it.


\section{Discussion}

\begin{figure}[H]
\centering
\includegraphics[width=1\columnwidth]{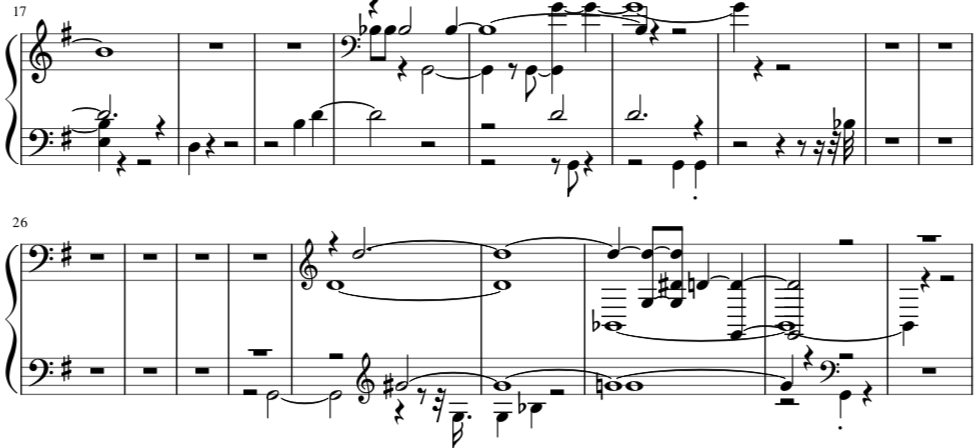}
\caption{MuseScore screenshot of measures 17 to 34 of \textit{sample 19} from the MAESTRO BPE experiment.}
\label{fig:maestro-bpe}
\end{figure}

In order to provide some qualitative insights from the generated content, we here present an individual subjective analysis of some of the results. Despite good results in terms of overall structureness presented in Table \ref{tab:maestro}, we noticed that on some occasions the model opts by resorting to rests (silence) for a few measures. This can obviously be a desirable outcome sometimes, but as is observable in Figure \ref{fig:maestro-bpe}, the rests from measure 24 to 28 seem to detract from the previous flow in terms of the musical idea behind it.

Furthermore, despite the frequentist approach in subword tokenization procedures, often giving emphasis to combinations of words/subwords that are more common in a given corpus, for the particular case of guitar-focused symbolic music generation with the DadaGP dataset, it is interesting to observe that tokens concerning guitar expressivity techniques are preserved, despite its diminished frequency when compared to note tokens.

\begin{figure}[H]
\centering
\includegraphics[width=1\columnwidth]{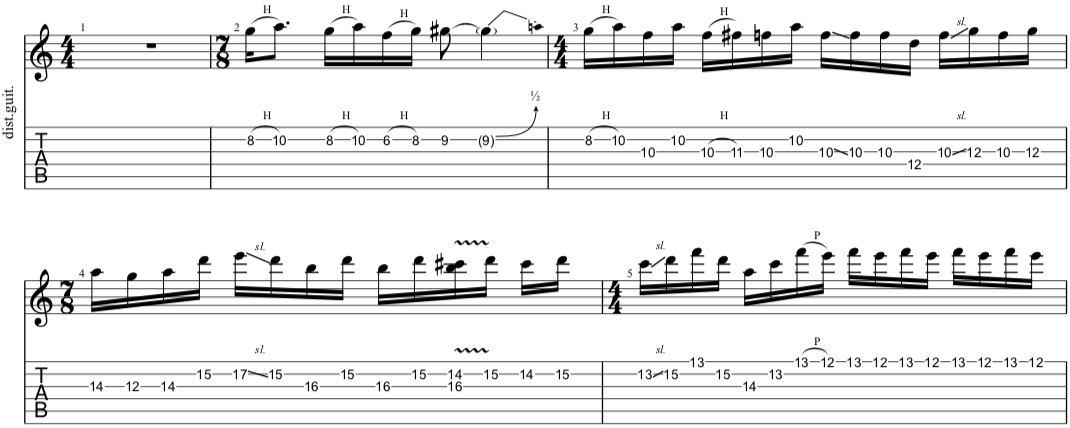}
\caption{GuitarPro screenshot of the first five measures from \textit{sample 18} from the DadaGP BPE experiment. Only the distorted guitar is visible.}
\label{fig:dadagp-bpe}
\end{figure}

As we can see from Figure \ref{fig:dadagp-bpe}, guitar-specific expressivity techniques such as hammer-ons and pull-offs, bends, slides, and vibrato, are adequately used. 

\begin{figure}[H]
\centering
\includegraphics[width=1\columnwidth]{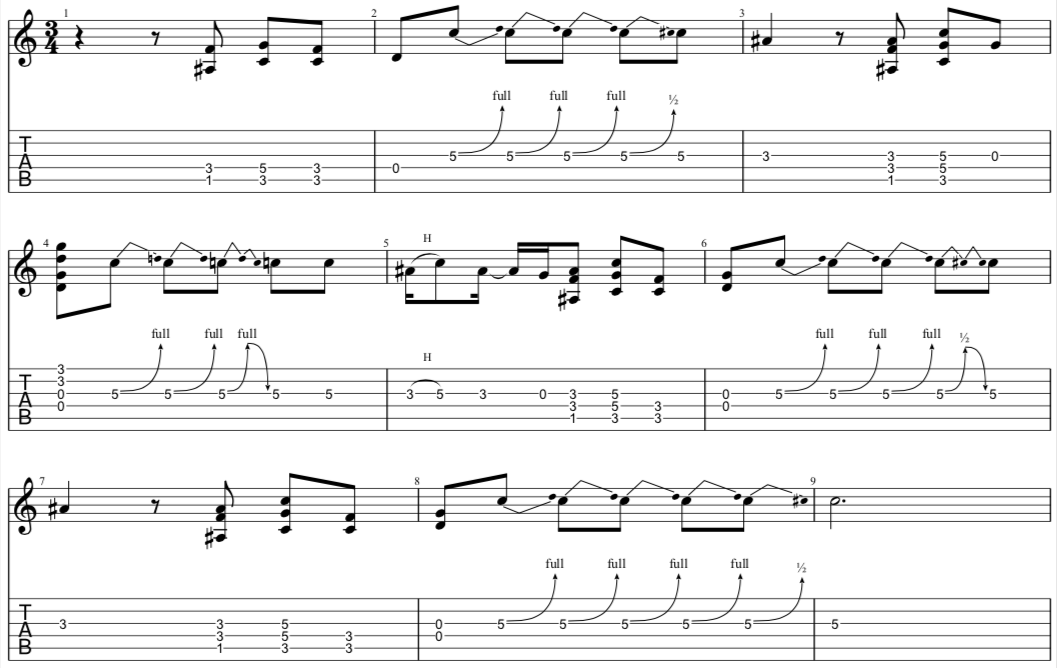}
\caption{GuitarPro screenshot of the first nine measures from \textit{sample 10} from the DadaGP Unigram experiment. Only the distorted guitar is visible.}
\label{fig:dadagp-bpe-struct}
\end{figure}

In order to visually support the arguments made in Section \ref{sec:objeval} towards the improvement in terms of `structureness' of the generated examples from the subword tokenization models, in Figure \ref{fig:dadagp-bpe-struct} we can observe that the model is able to call-back to the \textit{motifs} played on the first two measures (i.e. same pattern repeating in measures seven and eight). From measure 5, it is also interesting to observe that the model was able to refer to the same pattern that was introduced in the first two measures but also incorporated a few connecting notes in its first beats. 

\begin{figure}[H]
\centering
\includegraphics[width=1\columnwidth]{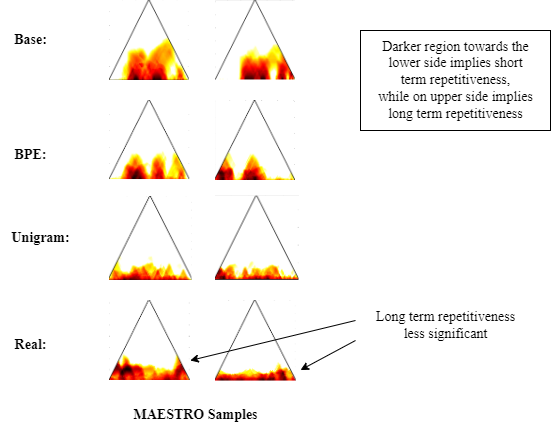}
\caption{Fitness scape plots for two generated samples from each of Base, BPE and Unigram along with real songs corresponding to the MAESTRO dataset. The x-axis represents the segment center (in sec) and y-axis represents the segment length (in sec) for any repetitive structure in music}
\label{fig:scapeplot}
\end{figure}

A similar observation on improvements in terms of structural representation can be made if we analyze the fitness scape plots for the generated songs against real songs. A sample of such observation is shown in Figure \ref{fig:scapeplot} for MAESTRO dataset samples. Contours in yellowish/brownish regions in scape plots represent the repetitive structure, with the lower side implying short-term and the higher side implying long-term repetitive structure. A better model is one that generates songs having a scape-plot similar to that of real songs since then the musical structure of the training dataset is captured in a more accurate way. As we can observe from the scape plots given in Figure \ref{fig:scapeplot}, there is a much less long-term repetitive structure in the real data. Both BPE and Unigram have their scape plots more similar to real songs, with lesser yellowish regions on the upper side of plots, than that of the base model. This suggests that the BPE and Unigram models capture the overall musical structure of real songs better than the base model, with Unigram being the best among the three. This observation is in coherence with Table \ref{tab:maestro}, suggesting improvements with the use of subword tokenization methods.    



\section{Conclusion and Future Work}

In this paper, we conducted an empirical study to assess the usefulness of subword tokenization methods in symbolic music generation. We objectively studied the change in performance of music generation models with the use of subword tokenization techniques such as BPE and Unigram while answering the two fundamental questions posed in Section \ref{sec:Intro}. Our study not only shows that BPE and Unigram, as data compression techniques, are not only able to represent the data in a more efficient manner but also improve the overall structure of music generated with model-inculcating subword tokenization. Furthermore, this result (and the trends in improvement) stands irrespective of the model, dataset, or musical tokenization used. Overall, from our study, we can conclude that the inclusion of subword tokenization in Symbolic Music generation has a potential impact on the performance of the model and structure of generated music, while allowing longer music to be generated at the same time.  

From this point onwards, future work can be multifaceted. One direction could be to explore the impact of vocabulary size on the model performance, i.e. how the vocal-performance trade-off. Similar to text-NLP, where we usually have a vocabulary size in the range of ~50k, we can increase the vocab size here as well to see how performance varies with changes in vocabulary size. Another interesting direction could be to explore whether the knowledge of Music theory can be leveraged in coordination with BPE or Unigram-type techniques, to develop a hybrid version of subword tokenization, involving concurrency (such as in Musical BPE) musical events with adjacency of tokens. In all, there is a large scope of future work that can be done, beyond this study.

\section*{Ethical Statement}
The training of large language models is an intensive computational process that requires vast amounts of energy, resulting in a significant carbon footprint. Cloud providers are increasingly offering services for training and hosting these models, but not all of them have committed to being carbon neutral, meaning they may contribute to greenhouse gas emissions. This is an important consideration when selecting a cloud provider for training models, as it has both environmental and ethical implications.

To minimize the impact of training large language models, one solution is to release pre-trained models. This approach enables others to use these models without having to undergo the energy-intensive process of training them from scratch. By releasing pre-trained models, we aim to reduce the carbon footprint associated with training and make it easier for others to use these models in a more sustainable manner. Additionally, releasing pre-trained models may also encourage collaboration and innovation in the field, further advancing the development of AI technology.

\section*{Acknowledgement} 
We would like to express our sincere gratitude to \href{https://www.ee.ntu.edu.tw/profile1.php?id=1090726}{Dr. Yi-Hsuan Yang}, for his supervision and guidance throughout this research project. Dr. Yang's expert advice and feedback were invaluable in shaping the direction and outcome of this study. His constant support and motivation were crucial in keeping us focused and inspired. We would also like to thank \href{http://www.facweb.iitkgp.ac.in/~sourav/}{Dr. Sourav Mukhopadhyay}, IIT Kharagpur for giving us an opportunity to work on this project as Adarsh's Master's Thesis Project.  


\bibliographystyle{named}
\bibliography{ijcai22}

\begin{thebibliography}{}

\bibitem[\protect\citeauthoryear{Conneau and Lample}{2019}]{Neurips}
Alexis Conneau and Guillaume Lample.
\newblock Cross-lingual language model pretraining.
\newblock In H.~Wallach, H.~Larochelle, A.~Beygelzimer, F.~d\textquotesingle
  Alch\'{e}-Buc, E.~Fox, and R.~Garnett, editors, {\em Advances in Neural
  Information Processing Systems}, volume~32. Curran Associates, Inc., 2019.

\bibitem[\protect\citeauthoryear{Dai \bgroup \em et al.\egroup }{2019a}]{Dai}
Zihang Dai, Zhilin Yang, Yiming Yang, Jaime Carbonell, Quoc~V Le, and Ruslan
  Salakhutdinov.
\newblock {Transformer-XL: {A}ttentive {L}anguage {M}odels {B}eyond a
  {F}ixed-{L}ength {C}ontext}.
\newblock In {\em Proc. of the 57th Annual Meeting of the Association for
  Computational Linguistics}, 2019.

\bibitem[\protect\citeauthoryear{Dai \bgroup \em et al.\egroup }{2019b}]{Tr-XL}
Zihang Dai, Zhilin Yang, Yiming Yang, Jaime Carbonell, Quoc~V. Le, and Ruslan
  Salakhutdinov.
\newblock Transformer-xl: Attentive language models beyond a fixed-length
  context, 2019.

\bibitem[\protect\citeauthoryear{Dai \bgroup \em et al.\egroup
  }{2022}]{failmusicrepete}
Shuqi Dai, Huiran Yu, and Roger~B. Dannenberg.
\newblock What is missing in deep music generation? a study of repetition and
  structure in popular music, 2022.

\bibitem[\protect\citeauthoryear{Devlin \bgroup \em et al.\egroup
  }{2018}]{BERT}
Jacob Devlin, Ming-Wei Chang, Kenton Lee, and Kristina Toutanova.
\newblock Bert: Pre-training of deep bidirectional transformers for language
  understanding, 2018.

\bibitem[\protect\citeauthoryear{Dong and Yang}{2018}]{Dong2018}
Hao-Wen Dong and Yi-Hsuan Yang.
\newblock {Convolutional Generative Adversarial Networks with Binary Neurons
  for Polyphonic Music Generation}.
\newblock In {\em Proc. of the 19th Int. Soc. for Music Information Retrieval
  Conf. (ISMIR)}, 2018.

\bibitem[\protect\citeauthoryear{Dong \bgroup \em et al.\egroup }{2020}]{Muspy}
Hao{-}Wen Dong, Ke~Chen, Julian~J. McAuley, and Taylor Berg{-}Kirkpatrick.
\newblock Muspy: {A} toolkit for symbolic music generation.
\newblock {\em CoRR}, abs/2008.01951, 2020.

\bibitem[\protect\citeauthoryear{Foote}{1999}]{ssm}
Jonathan Foote.
\newblock Visualizing music and audio using self-similarity.
\newblock In {\em Proceedings of the Seventh ACM International Conference on
  Multimedia (Part 1)}, MULTIMEDIA '99, page 77–80, New York, NY, USA, 1999.
  Association for Computing Machinery.

\bibitem[\protect\citeauthoryear{Fradet \bgroup \em et al.\egroup
  }{2021}]{Miditok}
Nathan Fradet, Jean-Pierre Briot, Fabien Chhel, Amal El~Fallah Seghrouchni, and
  Nicolas Gutowski.
\newblock Miditok: A python package for midi file tokenization.
\newblock In {\em Extended Abstracts for the Late-Breaking Demo Session of the
  22nd International Society for Music Information Retrieval Conference}, 2021.

\bibitem[\protect\citeauthoryear{Gage}{1994}]{Gage}
Philip Gage.
\newblock A new algorithm for data compression.
\newblock {\em The C Users Journal archive}, 12:23--38, 1994.

\bibitem[\protect\citeauthoryear{Gall{\'e}}{2019}]{galle-2019-investigating}
Matthias Gall{\'e}.
\newblock Investigating the effectiveness of {BPE}: The power of shorter
  sequences.
\newblock In {\em Proceedings of the 2019 Conference on Empirical Methods in
  Natural Language Processing and the 9th International Joint Conference on
  Natural Language Processing (EMNLP-IJCNLP)}, pages 1375--1381, Hong Kong,
  China, November 2019. Association for Computational Linguistics.

\bibitem[\protect\citeauthoryear{Guo \bgroup \em et al.\egroup }{2022}]{Infil}
Rui Guo, Ivor Simpson, Chris Kiefer, Thor Magnusson, and Dorien Herremans.
\newblock Musiac: An extensible generative framework for music infilling
  applications with multi-level control, 2022.

\bibitem[\protect\citeauthoryear{Hawthorne \bgroup \em et al.\egroup
  }{2019}]{MAESTRO}
Curtis Hawthorne, Andriy Stasyuk, Adam Roberts, Ian Simon, Cheng-Zhi~Anna
  Huang, Sander Dieleman, Erich Elsen, Jesse Engel, and Douglas Eck.
\newblock Enabling factorized piano music modeling and generation with the
  {MAESTRO} dataset.
\newblock In {\em International Conference on Learning Representations}, 2019.

\bibitem[\protect\citeauthoryear{Hiller}{2019}]{Hiller}
Lejaren Hiller.
\newblock {\em IV. Music Composed With Computers—A Historical Survey}, pages
  42--96.
\newblock Cornell University Press, Ithaca, NY, 2019.

\bibitem[\protect\citeauthoryear{Hsiao \bgroup \em et al.\egroup
  }{2021}]{Hsiao2021}
Wen-Yi Hsiao, Jen-Yu Liu, Yin-Cheng Yeh, and Yi-Hsuan Yang.
\newblock {{Compound Word Transformer}: {Learning} to Compose Full-Song Music
  Over Dynamic Directed Hypergraphs}.
\newblock In {\em Proc. of the AAAI Conf. on Artificial Intelligence}, 2021.

\bibitem[\protect\citeauthoryear{Huang and Yang}{2020}]{Huang2020}
Yu-Siang Huang and Yi-Hsuan Yang.
\newblock {Pop Music Transformer: Beat-based Modeling and Generation of
  Expressive Pop Piano Compositions}.
\newblock In {\em Proc. of the 28th ACM Int. Conf. on Multimedia}, 2020.

\bibitem[\protect\citeauthoryear{Huang \bgroup \em et al.\egroup
  }{2018}]{MusicTransformer}
Cheng{-}Zhi~Anna Huang, Ashish Vaswani, Jakob Uszkoreit, Noam Shazeer, Curtis
  Hawthorne, Andrew~M. Dai, Matthew~D. Hoffman, and Douglas Eck.
\newblock An improved relative self-attention mechanism for transformer with
  application to music generation.
\newblock {\em CoRR}, abs/1809.04281, 2018.

\bibitem[\protect\citeauthoryear{Huang \bgroup \em et al.\egroup
  }{2019}]{AnnaHuang2019}
Cheng-Zhi~Anna Huang, Ashish Vaswani, Jakob Uszkoreit, Noam Shazeer, Ian Simon,
  Curtis Hawthorne, Andrew~M. Dai, Matthew~D. Hoffman, Monica Dinculescu, and
  Douglas Eck.
\newblock {Music Transformer: Generating Music with Long-term Structure}.
\newblock In {\em Proc. of the 7th Int. Conf. on Learning Representations},
  2019.

\bibitem[\protect\citeauthoryear{Kudo and Richardson}{2018}]{SentencePiece}
Taku Kudo and John Richardson.
\newblock Sentencepiece: {A} simple and language independent subword tokenizer
  and detokenizer for neural text processing.
\newblock {\em CoRR}, abs/1808.06226, 2018.

\bibitem[\protect\citeauthoryear{Kudo}{2018}]{kudo-2018-subword}
Taku Kudo.
\newblock Subword regularization: Improving neural network translation models
  with multiple subword candidates.
\newblock In {\em Proceedings of the 56th Annual Meeting of the Association for
  Computational Linguistics (Volume 1: Long Papers)}, pages 66--75, Melbourne,
  Australia, July 2018. Association for Computational Linguistics.

\bibitem[\protect\citeauthoryear{Liu \bgroup \em et al.\egroup
  }{2019}]{Roberta}
Yinhan Liu, Myle Ott, Naman Goyal, Jingfei Du, Mandar Joshi, Danqi Chen, Omer
  Levy, Mike Lewis, Luke Zettlemoyer, and Veselin Stoyanov.
\newblock Roberta: {A} robustly optimized {BERT} pretraining approach.
\newblock {\em CoRR}, abs/1907.11692, 2019.

\bibitem[\protect\citeauthoryear{Liu \bgroup \em et al.\egroup
  }{2022}]{musicBPE}
Jiafeng Liu, Yuanliang Dong, Zehua Cheng, Xinran Zhang, Xiaobing Li, Feng Yu,
  and Maosong Sun.
\newblock Symphony generation with permutation invariant language model.
\newblock 2022.

\bibitem[\protect\citeauthoryear{Meade \bgroup \em et al.\egroup
  }{2019}]{Meade2019}
Nicholas Meade, Nicholas Barreyre, Scott~C Lowe, and Sageev Oore.
\newblock {Exploring Conditioning for Generative Music Systems with
  Human-Interpretable Controls}.
\newblock 2019.

\bibitem[\protect\citeauthoryear{Mielke \bgroup \em et al.\egroup
  }{2021}]{Subword}
Sabrina~J. Mielke, Zaid Alyafeai, Elizabeth Salesky, Colin Raffel, Manan Dey,
  Matthias Gallé, Arun Raja, Chenglei Si, Wilson~Y. Lee, Benoît Sagot, and
  Samson Tan.
\newblock Between words and characters: A brief history of open-vocabulary
  modeling and tokenization in nlp, 2021.

\bibitem[\protect\citeauthoryear{Muhamed \bgroup \em et al.\egroup
  }{2021}]{transformer-gan}
Aashiq Muhamed, Liang Li, Xingjian Shi, Suri Yaddanapudi, Wayne Chi, Dylan
  Jackson, Rahul Suresh, Zachary~C. Lipton, and Alexander~J. Smola.
\newblock Symbolic music generation with transformer-gans.
\newblock In {\em 35th AAAI Conference on Artificial Intelligence, {AAAI}
  2021}, 2021.

\bibitem[\protect\citeauthoryear{Müller \bgroup \em et al.\egroup
  }{2011}]{scape}
Meinard Müller, Peter Grosche, and Nanzhu Jiang.
\newblock A segment-based fitness measure for capturing repetitive structures
  of music recordings.
\newblock pages 615--620, 01 2011.

\bibitem[\protect\citeauthoryear{Neves \bgroup \em et al.\egroup
  }{2022}]{SentTransformerGAN}
Pedro Neves, Jose Fornari, and João Florindo.
\newblock Generating music with sentiment using transformer-gans, 2022.

\bibitem[\protect\citeauthoryear{Oore \bgroup \em et al.\egroup
  }{2018}]{midilike2018}
Sageev Oore, Ian Simon, Sander Dieleman, Douglas Eck, and Karen Simonyan.
\newblock This time with feeling: Learning expressive musical performance.
\newblock {\em Neural Computing and Applications}, 2018.

\bibitem[\protect\citeauthoryear{Park \bgroup \em et al.\egroup
  }{2020}]{park-etal-2020-empirical}
Kyubyong Park, Joohong Lee, Seongbo Jang, and Dawoon Jung.
\newblock An empirical study of tokenization strategies for various {K}orean
  {NLP} tasks.
\newblock In {\em Proceedings of the 1st Conference of the Asia-Pacific Chapter
  of the Association for Computational Linguistics and the 10th International
  Joint Conference on Natural Language Processing}, pages 133--142, Suzhou,
  China, December 2020. Association for Computational Linguistics.

\bibitem[\protect\citeauthoryear{Payne}{2019}]{christine_2019}
Christine Payne.
\newblock Musenet, 2019.

\bibitem[\protect\citeauthoryear{Peracha}{2020}]{nll_metrics}
Omar~A Peracha.
\newblock Improving polyphonic music models with feature-rich encoding.
\newblock 2020.

\bibitem[\protect\citeauthoryear{Radford \bgroup \em et al.\egroup
  }{2019}]{GPT}
Alec Radford, Jeff Wu, Rewon Child, David Luan, Dario Amodei, and Ilya
  Sutskever.
\newblock Language models are unsupervised multitask learners.
\newblock 2019.

\bibitem[\protect\citeauthoryear{Sarmento \bgroup \em et al.\egroup
  }{2021}]{Sarmento2021}
Pedro Sarmento, Adarsh Kumar, CJ~Carr, Zack Zukowski, Mathieu Barthet, and
  Yi-Hsuan Yang.
\newblock {DadaGP: a Dataset of Tokenized GuitarPro Songs for Sequence Models}.
\newblock In {\em Proc. of the 22nd Int. Soc. for Music Information Retrieval
  Conf.}, 2021.

\bibitem[\protect\citeauthoryear{Sarmento \bgroup \em et al.\egroup
  }{2023}]{gtrctrl}
Pedro Sarmento, Adarsh Kumar, Yu-Hua Chen, CJ~Carr, Zack Zukowski, and Mathieu
  Barthet.
\newblock Gtr-ctrl: Instrument and genre conditioning for guitar-focused music
  generation with transformers, 2023.

\bibitem[\protect\citeauthoryear{Sennrich \bgroup \em et al.\egroup
  }{2016}]{sennrich-etal-2016-neural}
Rico Sennrich, Barry Haddow, and Alexandra Birch.
\newblock Neural machine translation of rare words with subword units.
\newblock In {\em Proceedings of the 54th Annual Meeting of the Association for
  Computational Linguistics (Volume 1: Long Papers)}, pages 1715--1725, Berlin,
  Germany, August 2016. Association for Computational Linguistics.

\bibitem[\protect\citeauthoryear{Sturm \bgroup \em et al.\egroup
  }{2016}]{Sturm2016}
Bob~L. Sturm, Jo{\~{a}}o~Felipe Santos, Oded Ben-Tal, and Iryna Korshunova.
\newblock {Music transcription modelling and composition using deep learning}.
\newblock In {\em Proc. on the 1st Conf. on Computer Simulation of Musical
  Creativity}, 2016.

\bibitem[\protect\citeauthoryear{Tan and Herremans}{2020}]{Tan2020}
Hao~Hao Tan and Dorien Herremans.
\newblock {Music FaderNets: Controllable Music Generation Based On High-Level
  Features via Low-Level Feature Modelling}.
\newblock In {\em Proc. of the 21st Int. Soc. for Music Information Retrieval
  Conf.}, 2020.

\bibitem[\protect\citeauthoryear{Vaswani \bgroup \em et al.\egroup
  }{2017}]{Vaswani2017}
Ashish Vaswani, Noam Shazeer, Niki Parmar, Jakob Uszkoreit, Llion Jones,
  Aidan~N Gomez, {\L}ukasz Kaiser, and Illia Polosukhin.
\newblock {Attention Is All You Need}.
\newblock In {\em Proc. of the 31st Conf. on Neural Information Processing
  Systems}, 2017.

\bibitem[\protect\citeauthoryear{Wang \bgroup \em et al.\egroup
  }{2020}]{Wang2020}
Ziyu Wang, Dingsu Wang, Yixiao Zhang, and Gus Xia.
\newblock {Learning Interpretable Representation for Controllable Polyphonic
  Music Generation}.
\newblock In {\em Proc. of the 21st Int. Soc. for Music Information Retrieval
  Conf.}, Montr{\'{e}}al, Canada, 2020.

\bibitem[\protect\citeauthoryear{Wolf \bgroup \em et al.\egroup
  }{2020}]{wolf2020huggingfaces}
Thomas Wolf, Lysandre Debut, Victor Sanh, Julien Chaumond, Clement Delangue,
  Anthony Moi, Pierric Cistac, Tim Rault, Rémi Louf, Morgan Funtowicz, Joe
  Davison, Sam Shleifer, Patrick von Platen, Clara Ma, Yacine Jernite, Julien
  Plu, Canwen Xu, Teven~Le Scao, Sylvain Gugger, Mariama Drame, Quentin Lhoest,
  and Alexander~M. Rush.
\newblock Huggingface's transformers: State-of-the-art natural language
  processing, 2020.

\bibitem[\protect\citeauthoryear{Wu and Yang}{2020}]{Jazz}
Shih{-}Lun Wu and Yi{-}Hsuan Yang.
\newblock The jazz transformer on the front line: Exploring the shortcomings of
  ai-composed music through quantitative measures.
\newblock {\em CoRR}, abs/2008.01307, 2020.

\bibitem[\protect\citeauthoryear{Wu and Yang}{2022}]{composenembel}
Shih-Lun Wu and Yi-Hsuan Yang.
\newblock Compose \& embellish: Well-structured piano performance generation
  via a two-stage approach, 2022.

\bibitem[\protect\citeauthoryear{Wu \bgroup \em et al.\egroup
  }{2016}]{WordPiece}
Yonghui Wu, Mike Schuster, Zhifeng Chen, Quoc~V. Le, Mohammad Norouzi, Wolfgang
  Macherey, Maxim Krikun, Yuan Cao, Qin Gao, Klaus Macherey, Jeff Klingner,
  Apurva Shah, Melvin Johnson, Xiaobing Liu, Łukasz Kaiser, Stephan Gouws,
  Yoshikiyo Kato, Taku Kudo, Hideto Kazawa, Keith Stevens, George Kurian,
  Nishant Patil, Wei Wang, Cliff Young, Jason Smith, Jason Riesa, Alex Rudnick,
  Oriol Vinyals, Greg Corrado, Macduff Hughes, and Jeffrey Dean.
\newblock Google's neural machine translation system: Bridging the gap between
  human and machine translation, 2016.

\end{thebibliography}

\end{document}